\documentclass{ws-procs9x6-notrim}

\def\bfxi{\mbox{\boldmath $\xi$}}

\def\lQ{\Lambda_{\rm QCD}}

\newcommand{\nn}{\nonumber}
\newcommand{\be}{\begin{equation}}
\newcommand{\ee}{\end{equation}}
\newcommand{\bea}{\begin{eqnarray}}
\newcommand{\eea}{\end{eqnarray}}
\def\al{\alpha}
\def\als{\alpha_{\rm s}}
\def\siml{{\ \lower-1.2pt\vbox{\hbox{\rlap{$<$}\lower6pt\vbox{\hbox{$\sim$}}}}\ }} 
\def\simg{{\ \lower-1.2pt\vbox{\hbox{\rlap{$>$}\lower6pt\vbox{\hbox{$\sim$}}}}\ }}

\begin{document}

\title{Heavy Quarkonium Inclusive Decays: Theoretical Status and Perspectives}

\author{Antonio Vairo}

\address{Theory Division, CERN, 1211 Geneva 23, Switzerland\\ 
E-mail: antonio.vairo@cern.ch}

\maketitle

\abstracts{I review some recent progress and open problems in the calculation 
of heavy-quarkonium inclusive decay widths into light particles in the framework 
of QCD non-relativistic effective field theories.}

\section{Introduction}
In recent years, the study of heavy quarkonia ($b\bar{b}$, $c\bar{c}$, ...) 
has gone through several theoretical and experimental advances.

From the theoretical side the introduction of non-relativistic effective field 
theories (EFTs) of QCD\cite{nrqcd,Mont} has put our description of these systems
on the solid ground of QCD. It has made it possible, in the case of several
observables, to factorize the high-energy dynamics into matching coefficients 
calculable in perturbation theory and the non-perturbative QCD dynamics 
into few well-defined matrix elements 
to be fitted on the data or calculated on the lattice. Systematic
improvements are possible, either calculating higher-order corrections 
in the coupling constant or adding higher-order operators.

Heavy quarkonium, being a non-relativistic bound state, is characterized by
a hierarchy of energy scales $m$, $mv$ and $mv^2$, where $m$ is the heavy-quark 
mass and $v\ll 1$ the relative heavy-quark velocity.
A hierarchy of EFTs may be constructed by systematically integrating out 
modes associated to these energy scales. Integrating out degrees of freedom 
of energy $m$, which for heavy quarks can be done perturbatively, leads to  
non-relativistic QCD (NRQCD)\cite{nrqcd}. This EFT still contains the lower 
energy scales as dynamical degrees of freedom. In the last years, the problem 
of integrating out the remaining dynamical scales of NRQCD has been addressed 
by several groups and has now reached a solid level of conceptual
understanding (an extended list of references can be found in\cite{reveft}). 
The ultimate EFT obtained by subsequent matchings from QCD, where only the
lightest degrees of freedom of energy $mv^2$ are left dynamical, 
is called  potential NRQCD, pNRQCD\cite{Mont}. This EFT is close to a
quantum-mechanical description of the bound system and, therefore, as simple. 
It has been systematically explored in the dynamical
regime $\lQ \siml mv^2$ in\cite{long,logs} and in the regime $mv^2 \ll \lQ \siml mv$ 
in\cite{long,M12,sw}. The quantity $\lQ$ stands for the generic scale of 
non-perturbative physics.

From the experimental side new data have recently been produced 
for heavy-quarkonium observables. Measurements relevant to the
determination of heavy-quarkonium inclusive decay widths have come from 
Fermilab (E835)\cite{E835}, BES\cite{BES}, CLEO\cite{CLEO1,CLEO2} and Belle\cite{Belle}.

In the following I will review recent progress in our theoretical understanding
of inclusive and electromagnetic heavy-quarkonium decays. I will recall the
NRQCD factorization formulas in Sec. \ref{secnrqcd} and the pNRQCD
factorization in Sec. \ref{secpnrqcd}. The presented pNRQCD formulas apply 
to quarkonia that fulfil $mv \simg \lQ \gg mv^2$. In Sec. \ref{sechow}, I will 
discuss the main difficulties that still prevent us from obtaining precise 
predictions from the above formulas and indicate what progress has been made 
towards an eventual solution of these difficulties.

\section{NRQCD factorization}
\label{secnrqcd}
The NRQCD factorization formulas are obtained by separating contributions coming from 
degrees of freedom of energy $m$ from those coming from  degrees of freedom of lower 
energy. In the case of heavy-quarkonium decay widths,    
the first are encoded in the imaginary parts of the four-fermion
matching coefficients, $f,g_{1,8,ee,\gamma\gamma}(^{2S+1}L_J)$ and are ordered 
in powers of $\als(m)$. The second are encoded into the matrix elements 
of the four-fermion operators on the heavy-quarkonium states $|H\rangle$ 
($\langle \dots \rangle_{H} \equiv \langle H  | \dots  | H \rangle$). 
These are non-perturbative objects, which are counted, at least,  
in powers of $mv$. Matrix elements of higher dimensionality are suppressed in $v$.
Including up to the NRQCD four-fermion operators of dimension 8, the NRQCD factorization formulas 
for inclusive decay widths of heavy quarkonia into light hadrons ($LH$) read\cite{nrqcd,BBL0}:
\bea
&&\Gamma(V_Q (nS) \rightarrow LH) = {2\over m^2}\Bigg( 
{\rm Im\,}f_1(^3 S_1) \,  \langle O_1(^3S_1)\rangle_{V_Q(nS)}
\nn
\\
&&\quad
+ {\rm Im\,}f_8(^3 S_1)\, \langle O_8(^3S_1)\rangle_{V_Q(nS)}
+ {\rm Im\,}f_8(^1 S_0)\, \langle O_8(^1S_0)\rangle_{V_Q(nS)} 
\nn
\\
&&\quad 
+ {\rm Im\,}g_1(^3 S_1)\,
{\langle {\mathcal P}_1(^3S_1)\rangle_{V_Q(nS)} \over m^2}
+ {\rm Im\,}f_8(^3 P_0)\,
{\langle O_8(^3P_0)\rangle_{V_Q(nS)} \over m^2}
\nn
\\
&&\quad
+ {\rm Im\,}f_8(^3 P_1)\,
{\langle O_8(^3P_1)\rangle_{V_Q(nS)} \over m^2}
+ {\rm Im\,}f_8(^3 P_2)\,
{\langle O_8(^3P_2)\rangle_{V_Q(nS)} \over m^2}\Bigg),
\\
&&
\nn 
\\
&&\Gamma(P_Q (nS) \rightarrow LH) = {2\over m^2}\Bigg( 
{\rm Im\,}f_1(^1 S_0)\,   \langle O_1(^1S_0)\rangle_{P_Q(nS)}
\nn
\\
&&\quad
+ {\rm Im\,}f_8(^1 S_0)\, \langle O_8(^1S_0)\rangle_{P_Q(nS)}
+ {\rm Im\,}f_8(^3 S_1)\, \langle O_8(^3S_1)\rangle_{P_Q(nS)} 
\nn
\\
&&\quad 
+ {\rm Im\,}g_1(^1 S_0)\,
{\langle {\mathcal P}_1(^1S_0)\rangle_{P_Q(nS)} \over m^2}
+ {\rm Im\,}f_8(^1 P_1)\,
{\langle O_8(^1P_1)\rangle_{P_Q(nS)} \over m^2} \Bigg),
\\
&&
\nn
\\
&&\Gamma(\chi_Q(nJS)  \rightarrow LH)= 
{2\over m^2}\Bigg( {\rm Im \,}  f_1(^{2S+1}P_J)\, 
{\langle O_1(^{2S+1}P_J ) \rangle_{\chi_Q(nJS)} \over m^2}
\nn
\\
&&\quad
+ f_8(^{2S+1}S_S) \,\langle O_8(^1S_0 ) \rangle_{\chi_Q(nJS)} \Bigg).
\eea
At the same order the electromagnetic decay widths are given by:
\bea 
&&\Gamma(V_Q (nS) \rightarrow e^+e^-)= {2\over m^2}\Bigg( 
{\rm Im\,}f_{ee}(^3 S_1)\,   \langle O_{\rm EM}(^3S_1)\rangle_{V_Q(nS)}
\nn
\\
&&\quad
+ {\rm Im\,}g_{ee}(^3 S_1)\,
{\langle {\mathcal P}_{\rm EM}(^3S_1)\rangle_{V_Q(nS)} \over m^2}\Bigg),
\\
&&
\nn
\\
&&\Gamma(P_Q (nS) \rightarrow \gamma\gamma)= {2\over m^2}\Bigg( 
{\rm Im\,}f_{\gamma\gamma}(^1 S_0)\,   
\langle O_{\rm EM}(^1S_0)\rangle_{P_Q(nS)}
\nn
\\
&&\quad
+ {\rm Im\,}g_{\gamma\gamma}(^1 S_0)\,
{\langle {\mathcal P}_{\rm EM}(^1S_0)\rangle_{P_Q(nS)} \over m^2} \Bigg),
\\
&&
\nn
\\
&&\Gamma(\chi_Q(nJ1)  \rightarrow \gamma\gamma)= 
2 \, {\rm Im \,}  f_{\gamma\gamma}(^3P_J)\, 
{\langle O_{\rm EM}(^3P_J )\rangle_{\chi_Q(nJ1)}  \over m^4}, 
\quad J=0,2\,.
\label{chigg}
\eea
The symbols $V_Q$ and $P_Q$ indicate respectively the vector and pseudoscalar $S$-wave heavy 
quarkonium and the symbol $\chi_Q$ the generic $P$-wave quarkonium (the states
$\chi_Q(n10)$ and $\chi_Q(nJ1)$ are usually called $h_Q((n-1)P)$ and
$\chi_{QJ}((n-1)P)$, respectively). 

The operators $O,{\mathcal P}_{1,8,{\rm EM}}(^{2S+1}L_J)$ are the dimension $6$ and  $8$ 
four-fermion operators of the NRQCD Lagrangian. They are classified in dependence of their 
transformation properties under colour as singlets ($1$) and octets ($8$) and
under spin ($S$), orbital ($L$) and total angular momentum ($J$). 
The operators with the subscript EM are the singlet operators projected on 
the QCD vacuum. The explicit expressions of the operators may be found in\cite{nrqcd,sw}.

The imaginary parts of the four-fermion matching coefficients have been
calculated over the last twenty years to different levels of precision. 
In the following I will indicate, to the best of my knowledge, some recent 
literature where their updated value may be found. If the case arises, 
references to the original literature may also be found there.  
The imaginary parts of $f_8(^3S_1)$, $f_8(^1S_0)$,
$f_8(^3P_J)$, $f_1(^{2S+1}P_J)$, and $f_1(^1S_0)$ (this originally calculated
in\cite{BCGRswave}) have been calculated up to order $\als^3$ 
in\cite{Petrelli}. A different result for $f_1(^3P_0)$ and $f_1(^3P_2)$ is in\cite{BCGRpwave}.
The imaginary part of $f_1(^3S_1)$ has been calculated up to
order  $\als^4$ in\cite{ML}, the imaginary part of $g_1(^3S_1)$ at order
$\als^3$ may be found in\cite{GK}, the imaginary part of $g_1(^1S_0)$ at
order $\als^2$ in\cite{nrqcd} and the imaginary part of $f_8(^1P_1)$ at
order $\als^2$ in\cite{Maltoniphd}. Where the electromagnetic 
coefficients are concerned, the imaginary part of $f_{ee}(^3 S_1)$ has been calculated 
up to order $\al^2\als^2$ in\cite{lepdec}, the imaginary parts of 
$f_{\gamma\gamma}(^1 S_0)$ and $f_{\gamma\gamma}(^3 P_{0,2})$ up to order 
$\al^2\als$ (originally calculated in\cite{BCGRswave,BCGRpwave}) 
and $g_{ee}(^3 S_1)$ and  $g_{\gamma\gamma}(^1 S_0)$ up to order
$\al^2$ may be found in\cite{nrqcd}. 

The NRQCD matrix elements are poorly known. 
They may be fitted on the experimental decay data, as in\cite{Maltoni}, or
calculated on the lattice\cite{latbo}. The matrix elements of singlet operators may
be linked at leading order to the Schr\"odinger wave functions in the
origin\cite{nrqcd} and are often evaluated by means of potential models\cite{EichtenQuigg}.

\section{pNRQCD factorization}
\label{secpnrqcd}
The pNRQCD expressions of the NRQCD matrix elements are obtained by
integrating out degrees of freedom of energy larger than $mv^2$. 
Four different situations are possible: $\lQ \ll mv^2$,  $\lQ \sim
mv^2$,  $mv \gg \lQ \gg mv^2$ and $\lQ \sim mv$.
In the first situation the NRQCD matrix elements may be calculated
perturbatively up to non-perturbative corrections of the form
of local condensates.  This situation may apply to the ground state of 
bottomonium. Explicit formulas have been worked out for the electromagnetic 
decay of the $\Upsilon(1S)$ in\cite{TY}. In the situation $\lQ \sim
mv^2$ the NRQCD matrix elements turn out to be convolutions of 
non-local condensates and perturbative Green functions. In this situation,
which may apply to the ground states of bottomonium and charmonium,
only the real part of the spectrum has been studied\cite{logs}. 
Calculations of decay widths have not been done so far.
In this situation, however, the state and flavour-dependent part will 
not factor out from the integrals.
The situations $mv \gg \lQ \gg mv^2$ and $\lQ \sim mv$ have been studied 
in\cite{pw,sw}. In principle they may apply to all heavy quarkonium states 
below threshold and above the ground state. Both situations give the same 
physical results and will be summarized in the following.\footnote{ 
Possible threshold effects have been neglected. 
As suggested in\cite{puzzle} they may be large for the $\psi(2S)$.} The main point 
here is that, since $\lQ$ is well separated from the energy scale of the bound state,   
$mv^2$, the state dependence of the NRQCD matrix elements
factorizes in the quarkonium wave function, while contributions 
coming from excitations of order $\lQ$ are encoded into a few universal 
constants.

\subsection{pNRQCD factorization in the case $mv \! \simg \! \lQ \! \gg \! mv^2$}
The matrix element of the NRQCD operator $O$ on the 
heavy-quarkonium state $|H\rangle$ at rest, ${\bf P}=0$,
with quantum numbers $n$, $j$, $l$ and $s$, may be written in terms of the 
eigenstates $\!| \underbar{k}; {\bf x}_1 {\bf x}_2 \rangle\!$ of  $\!\!$
the NRQCD Hamiltonian as\cite{sw} 
\bea
\nn
&&\langle O \rangle_H =  
{1\over \langle {\bf P}=0| {\bf P}=0 \rangle }
\int \! d^3{r}\int \! d^3{r}'\int \! d^3{R}\int \! d^3{R}' \,
\langle {\bf P}=0| {\bf R}\rangle \langle njls |{\bf r}\rangle
\\
\nn
&& 
\qquad
\times 
\bigg[
\langle \underbar{0}; {\bf x}_1 {\bf x}_2|
\int \! d^3\xi \; O(\bfxi) 
\;
| \underbar{0}; {\bf x}_1^\prime {\bf x}_2^\prime \rangle
\bigg]
\langle {\bf R}'|{\bf P}=0\rangle \langle {\bf r}'|njls \rangle
\,,
\eea
where ${\bf r} = {\bf x}_1-{\bf x}_2$, ${\bf r}' = {\bf x}'_1-{\bf x}'_2$, 
${\bf R} = ({\bf x}_1+{\bf x}_2)/2$ and ${\bf R}' = ({\bf x}'_1+{\bf x}'_2)/2$.

In the situation $mv \simg \lQ \gg mv^2$ and assuming that all 
the higher gluonic excitations of the two heavy quarks develop a mass gap of 
order $\lQ$, the matrix element $\langle \underbar{0}; {\bf x}_1 {\bf x}_2|\, 
O(\bfxi) \,| \underbar{0}; {\bf x}_1^\prime {\bf x}_2^\prime
\rangle$ has been calculated order by order in $1/m$ in\cite{pw,sw}.
Once normalized to $m$, we get up to  ${\mathcal O}(v^3\times (\lQ^2/m^2,
E/m))$ for the NRQCD $S$-wave matrix elements and up to ${\mathcal O}(v^5)$ 
for the NRQCD $P$-wave matrix elements\cite{sw}:
\bea
\label{O13S1}
&&\langle O_1(^3S_1)\rangle_{V_Q(nS)}=
C_A {|R^V_{n0}({0})|^2 \over 2\pi}
\left(1-{E_{n0}^{(0)} \over m}{2{\mathcal E}_3 \over 9}
+{2{\mathcal E}^{(2,t)}_3 \over 3 m^2 }+{c_F^2{\mathcal B}_1 \over 3 m^2 }\right),
\\
&&\langle O_1(^1S_0)\rangle_{P_Q(nS)}=
C_A {|R^P_{n0}({0})|^2 \over 2\pi}
\left(1-{E_{n0}^{(0)} \over m}{2{\mathcal E}_3 \over 9}
+{2{\mathcal E}^{(2,t)}_3 \over 3 m^2}+{c_F^2{\mathcal B}_1 \over m^2}\right),
\\
&&\langle O_{\rm EM}(^3S_1)\rangle_{V_Q(nS)}=
C_A {|R^V_{n0}({0})|^2 \over 2\pi}
\left(1-{E_{n0}^{(0)} \over m}{2{\mathcal E}_3 \over 9}
+{2{\mathcal E}^{(2,{\rm EM})}_3 \over 3 m^2}+{c_F^2{\mathcal B}_1 \over 3 m^2}\right),\nn\\
\\
\label{OEM1S0}
&&\langle O_{\rm EM}(^1S_0)\rangle_{P_Q(nS)}=
C_A {|R^P_{n0}({0})|^2 \over 2\pi}
\left(1-{E_{n0}^{(0)} \over m}{2{\mathcal E}_3 \over 9}
+{2{\mathcal E}^{(2,{\rm EM})}_3 \over 3 m^2}+{c_F^2{\mathcal B}_1 \over m^2}\right),\nn\\
\\
&&
\langle O_1(^{2S+1}P_J ) \rangle_{\chi_Q(nJS)} = 
\langle O_{\rm EM}(^{2S+1}P_J ) \rangle_{\chi_Q(nJS)}  
={3 \over 2}{C_A \over \pi} |R^{(0)\,\prime}_{n1}({0})|^2,
\label{chio1}
\\
&&
\langle {\mathcal P}_1(^3S_1)\rangle_{V_Q(nS)}=
\langle {\mathcal P}_1(^1S_0)\rangle_{P_Q(nS)}=\langle {\mathcal P}_{\rm EM}(^3S_1)\rangle_{V_Q(nS)}
\nn\\
&&
\qquad\qquad\qquad
=\langle {\mathcal P}_{\rm EM}(^1S_0)\rangle_{P_Q(nS)}
=C_A {|R^{(0)}_{n0}({0})|^2 \over 2\pi}
\left(m E_{n0}^{(0)} -{\mathcal E}_1 \right),
\label{P13S1}
\\
&&
\langle O_8(^3S_1)\rangle_{V_Q(nS)}=
\langle O_8(^1S_0)\rangle_{P_Q(nS)}
\nn\\
&&\qquad\qquad\qquad\qquad\quad
=C_A {|R^{(0)}_{n0}({0})|^2 \over 2\pi}
\left(- {2 (C_A/2-C_f) {\mathcal E}^{(2)}_3 \over 3 m^2 }\right),
\\
&&
\langle O_8(^1S_0)\rangle_{V_Q(nS)}=
{\langle O_8(^3S_1)\rangle_{P_Q(nS)} \over 3}
\nn\\
&&\qquad\qquad\qquad\qquad\quad
=C_A {|R^{(0)}_{n0}({0})|^2 \over 2\pi}
\left(-{(C_A/2-C_f) c_F^2{\mathcal B}_1 \over 3 m^2 }\right),
\\
&&
\langle O_8(^3P_J) \rangle_{V_Q(nS)}=
{\langle O_8(^1P_1)\rangle_{P_Q(nS)} \over 3}
\nn\\
&&\qquad\qquad\qquad\qquad\quad
=(2J+1)\,C_A {|R^{(0)}_{n0}({0})|^2 \over 2\pi}
\left(-{(C_A/2-C_f) {\mathcal E}_1 \over 9 }\right),
\\
&&
\langle O_8(^1S_0)\rangle_{\chi_Q(nJS)} 
= {T_F\over 3}
{\vert R^{(0)\,\prime}_{n1}({0})\vert^2 \over \pi m^2} {\mathcal E}_3, 
\label{matoct}
\eea
where $R^V_{n0}=R_{n0}^{(0)}(1+{\mathcal O}(v))$ 
is the radial part of the vector $S$-wave function, 
$R^P_{n0}=R_{n0}^{(0)}(1+{\mathcal O}(v))$ 
the radial part of the pseudoscalar $S$-wave function, 
$R^{(0)\,\prime}_{n1}$ the derivative of the LO $P$-wave function 
and $E_{n0}^{(0)}\simeq M_H - 2m \sim m v^2$ the LO binding energy.
A consistent way to calculate the wave functions is from the real part 
of the pNRQCD Hamiltonian obtained up to order $1/m^2$ in\cite{M12}.
Any other $S$-wave dimension-6 matrix element is 0 at NNLO and any other
$S$-wave dimension-8 matrix element is 0 at LO. 
The symbol $c_F$ indicates the NRQCD chromomagnetic matching coefficient, which  
is known at NLO from\cite{ABN}.
All the non-perturbative dynamics is encoded into the wave functions and 
the six chromoelectric and chromomagnetic correlators ${\mathcal E}_1$, ${\mathcal E}_3$, 
${\mathcal E}^{(2)}_3$, ${\mathcal E}^{(2,t)}_3$, ${\mathcal E}^{(2,{\rm
 EM})}_3$ and ${\mathcal B}_1$. The exact definition of the correlators may be
found in\cite{sw}. It is important to note that, owing to the
factorization of the state and flavour dependence into the wave function, 
on the whole set of heavy-quarkonium states below threshold the number of
non-perturbative parameters needed to describe inclusive decays has diminished 
with respect to NRQCD, so that definite new predictions are possible.
In practice, one may consider ratios where the wave-function dependence drops
out\cite{pw,sw} and fix the correlators either through lattice calculations 
or through specific models of the QCD vacuum\cite{correlator} or on some set 
of experimental data. In particular, new predictions for $P$-wave
bottomonium inclusive decay-width ratios were made in this way in\cite{pw,pwphen} 
before the CLEO-III data were made available\cite{CLEO2}.

\section{How to use the above formulas}
\label{sechow}
Here I would like to point out that there are two potential problems that
may still prevent us from getting precise determinations of heavy-quar\-ko\-nium 
decay widths from the above formulas. These problems seem to be often underestimated 
in the existing literature. Eventually, they may lead to an underestimate  
of the errors associated to the calculated values of the decay widths or to 
the extracted values of the matrix elements and $\als(m)$.

$i)$ It has been discussed, in particular in\cite{mawang} (but see
also\cite{bope}), that higher-order operators, not considered in the above 
formulas, can be numerically quite relevant. This may be the case particularly 
for charmonium, where $v_c^2 \sim 0.3$, so that relativistic corrections 
are large, and for $P$-wave decays where the above formulas provide,
indeed, only the leading-order contribution in the velocity expansion.
In order to be specific, in the case of the $\chi_c$ decay width into two photons
the corrections of relative ${\mathcal O}(\lQ^2/m^2,E/m)$ to the formulas 
(\ref{chigg}) are given by\cite{mawang}:
\begin{eqnarray*}
{\delta \Gamma(\chi_{c0} \to \gamma \gamma) \over \Gamma^{(0)}(\chi_{c0} \to \gamma \gamma)}
&\simeq& -2.3 {E_{\chi_c}\over m_c} - 3 \, {\mathcal A}\left( {\lQ^2\over m^2_c}
  \right), \\
{\delta \Gamma(\chi_{c2} \to \gamma \gamma) \over \Gamma^{(0)}(\chi_{c2} \to \gamma \gamma)}
&\simeq& - 1.0  {E_{\chi_c}\over m_c} - 4 \, {\mathcal B}\left( {\lQ^2\over
      m^2_c} \right), 
\end{eqnarray*}
where ${\mathcal A}\left( {\lQ^2/ m^2_c}\right)$  and 
${\mathcal B}\left( {\lQ^2/ m^2_c}\right)$ are some specific combinations of matrix elements 
of order ${\lQ^2 / m^2_c}$. 
Choosing $E_{\chi_c}/m_c \simeq 0.3$ and $|{\mathcal A},{\mathcal B}(\lQ^2 /
m^2_c)| \simeq 0.1$
(or $0.3$)\footnote{The first choice corresponds to a hierarchy of the type 
$\lQ \sim mv^2$, the second of the type $\lQ \sim mv$.} 
the relative corrections may be $-(0.7\pm 0.3)$ (or $-(0.7\pm 0.9)$)
for $\Gamma(\chi_{c0} \to \gamma \gamma)$ and  $-(0.3\pm 0.4)$ (or $-(0.3\pm
1.2)$) for $\Gamma(\chi_{c2} \to \gamma \gamma)$ and, therefore, potentially 
as large as the leading one. 
The two $\pm$ values refer to the two possible choices of sign for ${\mathcal A}$ and
${\mathcal B}$. Note that for some specific choice of sign, large
cancellations may also occur.

The introduction of higher-order matrix elements may spoil the predictive
power of the NRQCD factorization. A way out is provided by pNRQCD, which can 
be used to reduce also these new matrix elements to a few universal correlators\cite{sw}.

$ii)$ The convergence of the perturbative series of the four-fermion matching coefficients 
is often bad. Let us consider, for instance, the following matching coefficients ($n_f = 3$):
\begin{eqnarray*}
&&
{\rm Im}f_{1}(^1S_0)  = (\dots)\times \left(1 + 11.1 \, {\als 
  \over \pi}\right), \\
&&
{\rm Im}f_{8}(^1S_0)  = (\dots)\times \left(1 + 13.7 \, {\als 
  \over \pi}\right), \\
&&
{\rm Im}f_{8}(^3S_1)  = (\dots)\times \left(1 + 10.3 \, {\als 
  \over \pi}\right), \\
&&
{\rm Im}f_{1}(^3P_0)  = (\dots)\times \left(1 + \left(13.6 - 0.44 \, \log {\mu\over 2m} \right)
\, {\als \over \pi}\right), \\
&&
{\rm Im}f_{1}(^3P_2)  = (\dots)\times \left(1 - \left(0.73  + 1.67 \, \log {\mu\over 2m} \right) 
\, {\als \over \pi}\right). 
\end{eqnarray*}
Apart from the case of ${\rm Im}f_{1}(^3P_2)$, the series in $\als$ of the other
coefficients does not seem to converge.
This behaviour may not be adjusted by a suitable choice of the factorization scale $\mu$, which enters
only in ${\rm Im}f_{1}(^3P_0)$; in this case lowering the factorization scale below
$2m$ would make the convergence worse.\footnote{In the case of the coefficient 
${\rm Im} f_{ee}(^3 S_1)$, which presents a similar problem, it has been noted 
by Beneke et al. in\cite{lepdec} that in the bottomonium case a suitable choice of the factorization 
scale may adjust the convergence of the series up to order $\al^2\als^2$.}
A solution may be provided by the resummation of the large contributions in the perturbative series 
coming from bubble-chain diagrams. This analysis has been successfully carried
out in some specific cases in\cite{chen}.
A general treatment is still missing, in particular in the case of $P$-wave decays.

Finally, I would like to note that neither contributions coming from higher-order matrix elements 
of the type described in $(i)$ nor resummations of large contributions in the perturbative series of the matching 
coefficients discussed in $(ii)$ have been considered in recent determinations 
of $\als$ (e.g. in\cite{ambrogiani,Maltoni}) and of the NRQCD matrix elements
(e.g. in\cite{Maltoni}) from the charmonium $P$-wave decay data.

\section{Conclusions}
The progress in our understanding of non-relativistic effective field theories
makes it possible to move beyond {\em ad hoc} phenomenological models and have 
a unified description of the different heavy-quarkonium observables, 
so that the same quantities determined from a set of data may be used in order to
describe other sets. Moreover, predictions based on non-relativistic EFTs 
are conceptually solid, and systematically improvable. Therefore, the recent
progress in the measurement of several heavy-quarkonium observables 
makes it meaningful to ask whether their precise theoretical determination 
is feasible. 

Here, I have focused on inclusive and electromagnetic heavy-quarkonium decays. 
In the framework of NRQCD, heavy-quarkonium decay widths may be expressed  
in terms of matrix elements, which, in the case of matrix elements of singlet 
operators, may be linked to the wave function in the origin. 
This is also the case of some matrix elements 
entering in the description of production processes\cite{nrqcd}.
In the framework of pNRQCD also some octet matrix elements may be expressed in terms of 
the wave function in the origin and some universal non-local gluon-field correlators. 
These correlators enter also in the expression of the masses of some 
heavy-quarkonium states\cite{logs,vill}. Specific predictions have been discussed 
in\cite{pw,sw}. There are, however, still some difficulties that hinder 
in several cases precise predictions of heavy-quarkonium decay observables.
These are essentially related to the control of higher-order corrections 
in the velocity and $\als$ expansion and have been addressed in the last section.
In principle, the tools to overcome these difficulties already exist, 
so that progress in the field is expected from the coordinated effort 
of the heavy-quarkonium community in the near future\cite{qwg}.

\section*{Acknowledgements}
I acknowledge the  support of the European Community through a Marie-Curie fellowship, 
contract No. HPMF-CT-2000-00733.

\end{document}